\documentclass[letterpaper, 10 pt, conference]{ieeeconf}  

\IEEEoverridecommandlockouts                              

\usepackage{amsfonts}
\usepackage{amsmath,amssymb,amsthm}
\usepackage{cite}
\usepackage{graphicx}
\usepackage{subfig}
\usepackage{mwe}
\usepackage{mathtools}
 \usepackage{scalefnt}
\usepackage{xcolor}
\usepackage[font=small]{caption}

\DeclarePairedDelimiterX{\norm}[1]{\lVert}{\rVert}{#1}
\allowdisplaybreaks
\newtheorem{theorem}{Theorem}
\newtheorem{remark}{Remark}
\newtheorem{proposition}{Proposition}
\newtheorem{lemma}{Lemma}

\title{\LARGE \bf
Safety Embedded Stochastic Optimal Control of Networked Multi-Agent Systems via Barrier States 
}

 \author{Lin Song$^{1}$, Pan Zhao$^{1}$, Neng Wan$^{1}$, and  Naira Hovakimyan$^{1}$
\thanks{*This work is  supported by Air Force Office of Scientific Research (AFSOR) (award \#FA9550-21-1-0411) and National Aeronautics and Space Administration (NASA) (awards \#80NSSC22M0070 and  \#80NSSC17M0051). }
   \thanks{$^{1}$Lin Song, Pan Zhao, Neng Wan, and  Naira Hovakimyan are with the Department of Mechanical Science and Engineering, University of Illinois at Urbana-Champaign, Urbana, IL 61801 USA {\tt\footnotesize \{linsong2, panzhao2, nengwan2, nhovakim\}@illinois.edu}}%
 }

\begin{document}

\maketitle
\thispagestyle{plain}
\pagestyle{plain}

\begin{abstract}
This paper presents a novel approach for achieving safe stochastic optimal control in networked multi-agent systems (MASs). The proposed method incorporates barrier states (BaSs) into the system dynamics to embed safety constraints. To accomplish this, the networked MAS is factorized into multiple subsystems, and each one is augmented with  BaSs for the central agent.  The optimal control law is obtained by solving the joint Hamilton-Jacobi-Bellman (HJB) equation on the augmented subsystem, which guarantees safety via the boundedness of the BaSs. The BaS-based optimal control technique yields safe control actions while maintaining optimality. The safe optimal control solution is approximated using path integrals. To validate the effectiveness of the proposed approach, numerical simulations are conducted on a cooperative UAV team in two different scenarios.

\end{abstract}

\section{INTRODUCTION}
 Optimal control has achieved remarkable success in both theory and applications~\cite{liberzon2011calculus,bryson2018applied}. Obtaining optimal control usually requires solving a nonlinear, second-order partial differential equation (PDE), known as Hamilton-Jacobi-Bellman (HJB) equation.  Stochastic optimal control (SOC) problems involve solving the control problem by minimizing expected costs~\cite{kappen2005linear}. By applying an exponential transformation to the value function~\cite{todorov2009efficient}, a linear-form HJB PDE is obtained, enabling related research including linearly-solvable optimal control (LSOC)~\cite{dvijotham2012linearly} and path-integral control (PIC)~\cite{pan2015sample,kappen2005linear}. The benefits of LSOC problems include compositionality~\cite{todorov2009compositionality,song2021compositionality} and the path-integral representation of the optimal control solution.  However, solving SOC problems in large-scale systems is challenging due to the curse of dimensionality~\cite{blondel2000survey}. To overcome computational challenges, many approximation-based approaches have been developed, such as path-integral (PI) formulation~\cite{theodorou2010generalized}, value function approximation~\cite{powell2011review}, and policy approximation~\cite{sutton1999policy}. In~\cite{van2008graphical}, a PI approach is used to approximate optimal control actions on multi-agent systems (MASs), and the optimal path distribution is predicted using the graphical model inference approach. A distributed PIC algorithm is proposed in~\cite{wan2021distributed}, in which a networked MAS is partitioned into multiple subsystems, and local optimal control actions are determined using local observations. However, these approaches seldom consider safety in the problem formulation, which may limit their real-world applications.

Safety refers to ensuring that a system's states remain within appropriate regions at all times for deterministic systems, or with a high probability for stochastic systems. Reachability analysis is a formal verification approach used to prove safety and performance guarantees for dynamical systems~\cite{summers2010verification,chapman2019risk}. Hamilton-Jacobi (HJ) reachability analysis identifies the initial states that the system needs to avoid as well as the associated optimal control for the sake of remaining safe~\cite{bansal2017hamilton}. However, computing the reachable set in reachability analysis is typically expensive, making it challenging to apply to multi-agent and high-dimensional systems. To enable safe optimal control, safety metrics can be incorporated into the optimal control framework, either as objectives or constraints. In~\cite{horowitz2014compositional}, temporal logic specifications are used as constraints for safety enforcement in optimal control development. The control barrier function (CBF) is a potent tool that can be used to enforce system safety by solving optimal control with constraints in a minimally invasive fashion~\cite{ames2016control}. CBF-based methods have also been extended to stochastic systems with high-probability guarantees~\cite{clark2021control,sarkar2020high,clark2019control}. A multi-agent CBF framework that generates collision-free controllers is discussed in~\cite{wang2017safety,borrmann2015control}. Furthermore, guaranteed safety-constraint satisfaction in the network system is achieved in~\cite{chen2020safety} under a valid assume-guarantee contract, with CBFs implemented onto subsystems. However, implementing CBFs as safety filters into the optimal control inputs may hinder ultimate optimality and be typically reactive to given constraints. Additionally, the feasibility of the quadratic programming (QP) introduced by CBF-based methods was not always guaranteed until the recent work in~\cite{xiao2022sufficient}. The barrier state (BaS) method is a novel methodology studied in~\cite{almubarak2021safety}, where the stability analysis of a BaS-augmented system encodes both stabilization and safety of the original system, and thus potential conflicts between control objectives and safety enforcement are avoided. In~\cite{almubarak2022discrete}, discrete BaS (DBaS) is employed with differential dynamic programming (DDP) in trajectory optimization, and it has been shown that bounded DBaS implies the generation of safe trajectories. The DBaSs have also been integrated into importance sampling to improve sample efficiency in safety-constrained sampling-based control problems in~\cite{gandhi2022safety}. 

Compared to CBF-based  methods that solve constrained optimization problems to determine certified-safe control actions, BaS-based safe control formulates the problem without explicit constraints; the safety notion is embedded in the solution boundedness, which prevents potential conflicts between control performance and safety requirements. However, the methodology of
addressing safety issues without sacrificing optimality in networked MASs remains an open problem. In this paper, we propose a safety-embedded SOC framework for networked MASs using BaSs proposed in~\cite{almubarak2021safety}. We adopt the MAS framework considered in~\cite{song2022generalization,wan2021cooperative}, where each agent computes optimal control based on local observations. However,~\cite{wan2021cooperative} does not consider system safety, while~\cite{song2022generalization} formulates safety concerns in the CBF framework and is potentially subject to the aforementioned issues.  To address the safety-guarantee deficiency in optimal controls, we augment the dynamics of the central agent in each subsystem with BaSs that embed safety constraints and formulate the optimal control problem using the augmented dynamics. 
Bounded solutions to the revised optimal control problem then ensure safety due to the characteristics of BaSs. 

The rest of the paper is structured as follows: Section~\ref{problem_form} introduces the preliminaries of formulating SOC problems and constructing BaS; Section~\ref{method} presents the safety-embedded SOC framework on MASs, along with the path integral formulation to approximate the control solution; and Section~\ref{simulation} provides numerical simulations in two scenarios to validate the effectiveness of the proposed approach. Finally, section~\ref{conclusion} concludes the paper and discusses future research directions. Several notations used in this paper are defined as follows: We use $|\mathcal{S}|$ denotes the cardinality of set $\mathcal{S}$, $\det(X)$ denotes the determinant of matrix $X$, $\textrm{tr}(X)$ denotes the trace of matrix $X$, $\nabla_{x}V$ and $\nabla_{xx}^2 V$ refer to the gradient and Hessian matrix of scalar-valued function $V$, and $\|v\|_M^2:= v^\top Mv$ denotes the weighted square norm.     

\section{PRELIMINARIES AND PROBLEM FORMULATION}\label{problem_form}
\subsection{Stochastic optimal control problems}
\subsubsection{MASs and factorial subsystems}\label{sec:distributed-network-framework}
\begin{figure}[!h]
\centering
\includegraphics[scale=0.25]{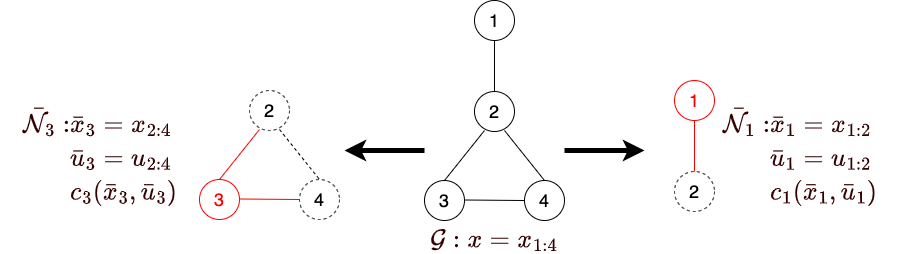}
\caption{MAS $\mathcal{G}$ and factorial subsystems $\bar{\mathcal{N}}_1$ and $\bar{\mathcal{N}}_3$.}
\label{fig_fact}
\end{figure}

We consider a MAS with $N$ homogeneous agents indexed by $\{1,2,\dots,N\}$. To describe the networked MAS, we use a connected and undirected graph $\mathcal{G} = \{\mathcal{V},\mathcal{E}\}$, where vertex $v_i \in \mathcal{V}$ represents agent $i$, and undirected edge $(v_i,v_j) \in \mathcal{E}$ indicates that agent $i$ and $j$ can communicate with each other. We define the index set of all agents neighboring agent $i$ as $\mathcal{N}_i$, and factorize the networked MAS into multiple subsystems $\bar{\mathcal{N}}_i = \mathcal{N}_i \cup \{i\}$, where each factorial subsystem consists of a central agent and all its neighboring agents. Figure~\ref{fig_fact} provides an illustrative example of the factorization scheme, where $\bar{x}_i$ and $\bar{u}_i$ denote the joint states and joint control actions  of factorial subsystem $\bar{\mathcal{N}}_i$.
 The local control action $u_j$ is determined by minimizing a joint cost function on subsystem $\bar{\mathcal{N}}_j$, which depends on the local observation  $\bar{x}_j$. Computing optimal control actions and sampling are therefore related to the size of each subsystem, rather than the entire network, which reduces computational complexity. For more discussions on the distributed control for LSOC problems on MASs, interested readers can refer to~\cite{wan2021distributed}.

\subsubsection{Stochastic optimal control of MASs}
We use the It$\hat{\textrm{o}}$ diffusion process to describe the joint dynamics of subsystem $\bar{\mathcal{N}}_i$ in a networked MAS consisting of $N$ homogeneous agents governed by mutually independent passive dynamics. The process is represented by the following equation: 
\begin{equation}\label{eq:cont_dyn_mas}
    d\bar{x}_i = \bar{g}_i(\bar{x}_i,t)dt+\bar{B}_i(\bar{x}_i)[\bar{u}_i(\bar{x}_i,t)dt+\bar{\sigma}_i d\bar{\omega}_i], 
\end{equation}
where  $\bar{x}_i = [x_i^\top,x_{j\in \mathcal{N}_i}^\top]^\top \in \mathbb{R}^{M \cdot |\bar{\mathcal{N}}_i|}$ is the joint state vector and $M$ represents the state dimension of each individual agent,  $\bar{g}_i(\bar{x}_i,t) = [g_i(x_i,t)^\top,g_{j\in \mathcal{N}_i}(x_j,t)^\top]^\top \in \mathbb{R}^{M\cdot|\bar{\mathcal{N}}_i|}$ represents the joint passive dynamics, which includes the passive dynamics of the individual agent $i$ and its neighbors $j \in \mathcal{N}_i$. $\bar{B}_i(\bar{x}_i) = \textrm{diag}\{B_i(x_i), B_{j \in \mathcal{N}_i}(x_j)\} \in \mathbb{R}^{M \cdot |\bar{\mathcal{N}}_i|\times P \cdot |\bar{\mathcal{N}}_i|}$ is the joint control matrix, $\bar{u}_i(\bar{x}_i,t) = [u_i(\bar{x}_i,t)^\top,u_{j\in \mathcal{N}_i}(\bar{x}_i,t)^\top]^\top \in \mathbb{R}^{P\cdot |\bar{\mathcal{N}}_i|}$ is the joint control action vector, and $\bar{\omega}_i = [\omega_i^\top,\omega_{j \in \mathcal{N}_i}^\top]^\top \in \mathbb{R}^{P \cdot |\bar{\mathcal{N}}_i|}$ is the joint noise vector with covariance matrix $\bar{\sigma}_i = \textrm{diag}\{\sigma_i,\sigma_{j\in \mathcal{N}_i}\} \in \mathbb{R}^{P \cdot |\bar{\mathcal{N}}_i| \times P \cdot |\bar{\mathcal{N}}_i|}$. To ensure the uniqueness of the solution, we assume that $\bar{g}_i, \bar{B}_i, \bar{\sigma}_i$ are locally Lipschitz continuous.

We use $\bar{\mathcal{B}}_i$ to denote the set of joint terminal states and $\bar{\mathcal{I}}_i$ to denote the set of joint non-terminal states. The entire allowable joint state space $\bar{\mathcal{S}}_i$ is partitioned into $\bar{\mathcal{I}}_i$ and $\bar{\mathcal{B}}_i$. For $\bar{x}_i \in \bar{\mathcal{I}}_i$, we define the running cost function as
\begin{equation*}
    c_i(\bar{x}_i,\bar{u}_i) = q_i(\bar{x}_i) + \frac{1}{2}\bar{u}_i(\bar{x}_i,t)^\top \bar{R}_i\bar{u}_i(\bar{x}_i,t),
\end{equation*}where $q_i(\bar{x}_i) \in \mathbb{R}_{\ge 0}$ is a joint state cost, and $\bar{u}_i(\bar{x}_i,t)^\top \bar{R}_i\bar{u}_i(\bar{x}_i,t)$ is a control-quadratic cost with positive definite matrix $\bar{R}_i \in \mathbb{R}^{P \cdot |\bar{\mathcal{N}}_i| \times P \cdot |\bar{\mathcal{N}}_i|}$. When $\bar{x}_i^{t_f} \in \bar{\mathcal{B}}_i$, the terminal cost function is denoted by $\phi_i(\bar{x}_i^{t_f})$, where $t_f$ is the final time. We also have the terminal cost function $\phi_i(x_i^{t_f})$ defined for $x_i^{t_f} \in \mathcal{B}_i$. In the first exit formulation, $t_f$ is determined online as the first time a joint state $\bar{x} \in \bar{\mathcal{B}}_i$ is reached. The cost-to-go function $J^{\bar{u}_i}(\bar{x}_i^t,t)$ under joint control action $\bar{u}_i$ is defined as \begin{equation}\label{eq:sas-cost-to-go}
    J^{\bar{u}_i}(\bar{x}_i^t,t) = \mathbb{E}_{\bar{x}_i^t,t}^{\bar{u}_i}[\phi_i(\bar{x}_i^{t_f})+\int_t^{t_f}c_i(\bar{x}_i(\tau),\bar{u}_i(\tau))d\tau],
\end{equation}
where the expectation is taken with respect to the probability measure under which $\bar{x}_i$ satisfies~\eqref{eq:cont_dyn_mas} under given joint control $\bar{u}_i$ starting from the initial condition $\bar{x}_i^t$. The optimal cost-to-go function (or value function) is formulated as \begin{equation*}
    V_i(\bar{x}_i^t,t) = \min_{\bar{u}_i}  J^{\bar{u}_i}(\bar{x}_i^t,t),
\end{equation*}which is the minimum expected cumulative running cost starting from joint state $\bar{x}_i^t$. For the sake of brevity, we use the notation $\bar{x}_i$ to represent $\bar{x}_i(t)$ and $\bar{x}_i^t$ in the following context.  

Facilitated by the exponential transformation of the value function, the optimal control action for the stochastic system~\eqref{eq:cont_dyn_mas} can be expressed in a linear form. The linear-form optimal control solution was initially proposed for a single-agent system in~\cite{fleming1977exit}, and later extended to a multi-agent scenario in~\cite{wan2021cooperative}. Here, we summarize the main results. 
The desirability function $Z(\bar{x}_i,t) = \exp[-V_i(\bar{x}_i,t)/\lambda_i]$ is defined over the joint state $\bar{x}_i$,
and the nonlinearity cancellation condition  $\bar{\sigma}_i\bar{\sigma}_i^\top = \lambda_i\bar{R}_i^{-1}$ is imposed to eliminate the nonlinear terms. Then, the linear-form joint optimal control action for the factorial subsystem $\bar{\mathcal{N}}_i$ in the networked MAS under the discussed decentralization topology takes the form of\begin{equation}\label{eq:mas-optimal-control-pf}\small
    \bar{u}_i^*(\bar{x}_i,t) = \bar{\sigma}_i\bar{\sigma}_i^\top \bar{B}_i(\bar{x}_i)^\top\nabla_{\bar{x}_i} Z(\bar{x}_i,t)/Z(\bar{x}_i,t). 
\end{equation}The nonlinearity cancellation condition essentially indicates that control is costly when the noise variance in a control channel is low, and therefore large control effort is avoided. 
\begin{remark}\label{remark_central}
In factorial subsystem $\bar{\mathcal{N}}_i$, we only obtain the local optimal control action $u_i^*(\bar{x}_i,t)$ for the central agent $i$, and local optimal control actions for the non-central agents $j (j \in \mathcal{N}_i)$ in $\&{\mathcal{N}}_i$ are computed from subsystem $\bar{\mathcal{N}}_j$.
\end{remark}

\subsection{Safety-embedded control via BaSs}\label{sec:prelim_bas}
The recently proposed safety-embedded control methodology through BaSs is a novel approach  that enforces CBF constraints by augmenting the original system with BaSs~\cite{almubarak2021safety,almubarak2022discrete,almubarak2022robust}. This methodology has demonstrated that the boundedness of the BaSs implies the safety of the original system. In this context, we provide a brief summary of the construction of BaS for nonlinear control-affine systems, which our proposed optimal control framework on the augmented MASs embedding safety builds upon. We consider nonlinear control-affine dynamical systems modeled as
\begin{equation}\label{eq:pf-ctrl-affine-dynamics}
    \dot x = g(x) + b(x)u,
\end{equation}
where $x \in \mathcal{D} \subset \mathbb{R}^M$, $u \in \mathbb{R}^P$, $g:\mathbb{R}^M \to \mathbb{R}^M$, and $b: \mathbb{R}^M \to \mathbb{R}^{M \times P}$ are continuously differentiable functions. The set $\mathcal{D}$ is the domain of operation, and we denote $h: \mathcal{D} \to \mathbb{R}$ as a continuously differentiable function representing the safe set, e.g., the safe operating region $\mathcal{C} := \{x \in \mathcal{D}|h(x) > 0\}$. A critical property of the scalar-valued barrier function (BF) is that its value remains bounded except when $x$ approaches the boundaries of the safe operating region $\mathcal{C}$. We consider the composite barrier function (BF) in the form of $\beta(x) = B \circ h(x)$, where  $h(x)$ defines a safe set. The dynamics of the BaS $z$ are modeled as
\begin{equation}\label{eq:pf-bas-nonlinear-sys}
\small
    \dot z = \phi_0(z + \beta_0)\dot h(x) -\gamma \phi_1(z + \beta_0, h(x)),
\end{equation}
where $\beta_0 = \beta(0), \dot{h}(x) = L_g h(x) + L_b h(x) u, \gamma \in \mathbb{R}_{> 0}$, $\phi_0(\cdot), \phi_1(\cdot)$ are analytic functions formulated based on the choice of BF $B(\cdot)$, $h(x)$, and are subject to certain conditions proposed in~\cite{almubarak2021safety}. We introduce a lemma from~\cite{almubarak2021safety} that connects the boundedness of the BaSs with the safety of the generated trajectories.
\begin{lemma}[\cite{almubarak2021safety}]\label{lemma1}
Suppose that $z(0) = \beta(x(0)) - \beta(0)$ and $\beta(x(0)) < \infty$, the BaS $z(t)$ generated from~\eqref{eq:pf-bas-nonlinear-sys} with the nonlinear dynamics~\eqref{eq:pf-ctrl-affine-dynamics} is bounded if and only if $\beta(x(t))$ is bounded for all $t > 0$.
\end{lemma}
\begin{remark}\label{remark_bas}
From Lemma~\ref{lemma1}, we can observe that ensuring the boundedness of the BaS implies boundedness of $\beta(x(t))$. This implies that $h(x(t)) > 0$ due to the properties of the BF, ensuring that the system trajectories always remain in the safe operating region $\mathcal{C} = \{x \in \mathcal{D}|h(x) > 0\}$.
\end{remark}

\section{BARRIER STATE AUGMENTED STOCHASTIC OPTIMAL CONTROL OF MASs}\label{method}
\subsection{Safe optimal control of MASs via BaSs}~\label{sec:aug_opt_ctrl_mas}
Using the system factorization methodology introduced in Section~\ref{sec:distributed-network-framework}, we can now formulate a safety-embedded optimal control on MASs using BaSs. Based on Remark~\ref{remark_central} and the distributed control framework we build upon, we conjecture that appending only the BaSs corresponding to the central agent for each subsystem is sufficient since only agent $i$ samples the local optimal control action from the computation results on subsystem $\bar{\mathcal{N}}_i$. By determining a bounded solution to the optimal control problem on each augmented subsystem, we can certify the safety of the included central agent. Taking certified-safe control actions for each agent from its corresponding subsystem (in which it acts as the central agent) collectively establishes the safety of the entire network. 

We first consider the continuous-time dynamics of  agent $i$ within $\bar{\mathcal{N}}_i$, which can be described by the following It$\hat{\textrm{o}}$ process\begin{equation}\label{eq:sas-original-dyn}
\small
    dx_i = g_i(x_i,t)dt + B_i(x_i)[u_i(\bar{x}_i,t)dt + \sigma_i d\omega_i],
\end{equation}where $x_i \in \mathbb{R}^M$ is the state vector, $g_i(x_i,t) \in \mathbb{R}^M$ is the passive dynamics vector, $B_i(x_i)\in \mathbb{R}^{M \times P}$ is the control matrix, $u_i(\&x_i,t) \in \mathbb{R}^P$ is the control action vector, and $\omega_i \in \mathbb{R}^{P}$ is the noise vector with covariance matrix $\sigma_i \in \mathbb{R}^{P  \times P}$. The construction of the BaS introduced in Section~\ref{sec:prelim_bas} is based on general nonlinear control-affine dynamical systems. Here, we specifically consider the dynamics in form of~\eqref{eq:sas-original-dyn} and construct the corresponding BaSs. 

We use $N_s$ independent BaSs to describe all $N_s$ safety constraints of interest for each agent. For agent $i$, we denote the function modeling $j$-th constraint as $h_j(x_i)$ and the corresponding BaS associated with $h_j(x_i)$ as $z_{i(j)}$. Suppose we use the inverse BF, i.e., $\beta(x_i) = 1/h(x_i)$, and we choose $\phi_0(\xi) = -\xi^2$ and $\phi_1(\xi,\eta)=\eta \xi^2-\xi$ as in~\cite{almubarak2021safety}. Then, the dynamics for each single BaS are reduced to:
\par\nobreak
\vspace{-4mm}
{\small
  \setlength{\abovedisplayskip}{6pt}
  \setlength{\belowdisplayskip}{\abovedisplayskip}
  \setlength{\abovedisplayshortskip}{0pt}
  \setlength{\belowdisplayshortskip}{3pt}
  \begin{align}\label{eq:sas_single_bas_dynamics_simplified}
     &dz_{i(j)} = \nonumber \\
    & [-(z_{i(j)} + \beta_{0(j)})^2(\frac{\partial h_j}{\partial x_i} g_i(x_i,t) + \gamma h_j(x_i)) + \gamma (z_{i(j)}+\beta_{0(j)})]dt \nonumber \\
    &- (z_{i(j)} + \beta_{0(j)})^2 \frac{\partial h_j}{\partial x_i} B_i(x_i)[u_i(\&x_i,t) dt+\sigma_i d\omega_i] \nonumber \\
    &:= g_{bi(j)}(x_i,z_{i(j)},t)dt + B_{bi(j)}(x_i,z_{i(j)})[u_i(\&x_i,t)dt + \sigma_i d\omega_i], 
  \end{align}
}with $\beta_{0(j)} = 1/h_j(0)$, $z_{i(j)} \in \mathbb{R}$, $g_{bi(j)}(x_i,z_{i(j)},t) \in \mathbb{R}$, $B_{bi(j)}(x_i,z_{i(j)}) \in \mathbb{R}^{1 \times P}$, and $u_i(\&x_i,t) \in \mathbb{R}^P$.  Since each agent is subject to $N_s$ constraints, which are modeled by the BaS dynamics in~\eqref{eq:sas_single_bas_dynamics_simplified}, we can represent the reorganized BaS dynamics for agent $i$ as:
\begin{equation}\label{eq:sas_multi_bas_dynamics}
   \small
   dz_i = g_{bi}(x_i,z_i,t)dt + B_{bi}(x_i,z_i)[u_i(\&x_i,t)dt + \sigma_i d\omega_i],
\end{equation}
with  $z_i = [z_{i(1)},\dots,z_{i(N_s)}]^\top \in \mathbb{R}^{N_s}$, $g_{bi}(x_i,z_i,t) = [g_{bi(1)}(x_i,z_i,t),\dots,g_{bi(N_s)}(x_i,z_i,t)]^\top \in \mathbb{R}^{N_s}$, $B_{bi}(x_i,z_i) = [B_{bi(1)}(x_i,z_i)^\top,\dots,B_{bi(N_s)}(x_i,z_i)^\top]^\top \in \mathbb{R}^{N_s \times P}$, and the individual elements $z_{i(j)}$, $g_{bi(j)}(x_i,z_i,t)$, and $B_{bi(j)}(x_i,z_i)$ are defined by~\eqref{eq:sas_single_bas_dynamics_simplified}.
Next, we integrate the BaS dynamics~\eqref{eq:sas_multi_bas_dynamics} corresponding to all $N_s$ constraints that the central agent $i$ is subject to. As a result, the augmented joint dynamics take the form of
\begin{equation}\label{eq:mas_aug_dynamics}
    d\bar{Y}_i = \bar{g}_i^\dagger (\bar{Y}_i,t)dt + \bar{B}_i^\dagger (\bar{Y}_i) [\bar{u}_i(\bar{Y}_i,t) dt + \bar{\sigma}_id\bar{\omega}_i],
\end{equation}
where  $\bar{Y}_i = [x_i^\top,z_i^\top,x_{j \in \mathcal{N}_i}^\top]^\top := [Y_i^\top, x_{j \in \mathcal{N}_i}^\top]^\top \in \mathbb{R}^{M\cdot |\bar{\mathcal{N}}_i| + N_s}$ is the augmented joint state vector,  $\bar{g}_i^\dagger (\bar{Y}_i,t) = [g_i(x_i,t)^\top, g_{bi}(x_i,z_i,t)^\top, g_{j \in \mathcal{N}_i}(x_j,t)^\top]^\top := [g_i^\dagger (Y_i,t)^\top, g_{j \in \mathcal{N}_i}(x_j,t)^\top]^\top \in \mathbb{R}^{M\cdot |\bar{\mathcal{N}}_i| + N_s}$ is the augmented joint passive dynamics vector, $\bar{B}_i^\dagger (\bar{Y}_i) = \textrm{diag}\{B_i^\dagger (Y_i),B_{j \in \mathcal{N}_i}(x_j)\} \in \mathbb{R}^{(M\cdot |\bar{\mathcal{N}}_i| + N_s) \times (P \cdot |\bar{\mathcal{N}}_i|)}$ with $B_i^\dagger (Y_i) = [B_i(x_i)^\top, B_{bi}(x_i,z_i)^\top]^\top \in \mathbb{R}^{(M+N_s) \times P}$ is the augmented joint control matrix, $\bar{u}_i(\bar{Y}_i,t) = [u_i(\bar{x}_i,t)^\top, u_{j \in \mathcal{N}_i}(\bar{x}_i,t)^\top]^\top \in \mathbb{R}^{P \cdot |\bar{\mathcal{N}}_i|}$ is the joint control action for the augmented system, $\bar{\omega}_i = [\omega_i^\top, \omega_{j \in \mathcal{N}_i}^\top]^\top \in \mathbb{R}^{P \cdot |\bar{\mathcal{N}}_i|}$ is the joint noise vector, and $\bar{\sigma}_i = \textrm{diag}\{\sigma_i, \sigma_{j \in \mathcal{N}_i}\} \in \mathbb{R}^{(P \cdot |\bar{\mathcal{N}}_i|) \times  (P \cdot |\bar{\mathcal{N}}_i|)}$ is the covariance matrix of noise vector  $\bar{\omega}_i$. In particular, the BaS vector for agent $i$ is denoted by $z_i = [z_{i(1)},z_{i(2)},\dots,z_{i(N_s)}]^\top \in \mathbb{R}^{N_s}$, where each $z_{i(j)}$ is subject to the BaS dynamics in~\eqref{eq:sas_single_bas_dynamics_simplified}. Since only some states are directly actuated, we can partition the augmented joint state $\bar{Y}_i$ into directly actuated states $\bar{Y}_{i(d)} \in \mathbb{R}^{D \cdot |\bar{\mathcal{N}}_i| + N_{SD}}$ and non-directly actuated states $\bar{Y}_{i(n)} \in \mathbb{R}^{U \cdot |\bar{\mathcal{N}}_i| + N_{SU}}$. Then, we can express $\bar{Y}_i = [\bar{Y}_{i(n)}^\top, \bar{Y}_{i(d)}^\top]^\top$, where $U$ and $D$ denote the dimensions of non-directly and directly actuated states for one agent, and $N_{SU}$, $N_{SD}$ denote the dimension of non-directly and directly actuated BaSs for one agent. Using this notation, we can rewrite the augmented joint dynamics in~\eqref{eq:mas_aug_dynamics} in the following partitioned vector form\begin{equation}\label{eq:mas_aug_dynamics_partitioned}
    \setlength\arraycolsep{1pt}
{\scalefont{0.65}\begin{bmatrix}
d\bar{Y}_{i(n)}\\
d\bar{Y}_{i(d)}
\end{bmatrix} = \begin{bmatrix}
 \bar{g}^\dagger_{i(n)}(\bar{Y}_i,t)\\
 \bar{g}^\dagger_{i(d)}(\bar{Y}_i,t)
\end{bmatrix}dt
+ \begin{bmatrix}
\mathbf{0}\\
 \bar{B}^\dagger_{i(d)}(\bar{Y}_i)
 \end{bmatrix}\big[\bar{u}_i(\bar{Y}_i,t) dt + \bar{\sigma}_id\bar{\omega}_i\big] , }
\end{equation}
where $\mathbf{0}$ denotes a zero matrix with the appropriate dimension. We define the joint running cost function  involving the augmented joint state $\&Y_i$ of the network $\bar{\mathcal{N}}_i$ as
\begin{equation}\label{eq:mas-aug-immediate-cost} \small
    c_i(\bar{Y}_i,\bar{u}_i) = q_i(\bar{Y}_i) + \frac{1}{2}\bar{u}_i(\bar{Y}_i,t)^\top \bar{R}_i \bar{u}_i(\bar{Y}_i,t),
\end{equation}
where $\bar{R}_i \in \mathbb{R}^{P \cdot \bar{\mathcal{N}}_i \times P \cdot \bar{\mathcal{N}}_i}$ is positive definite. Here, we assume that the control weights of each agent are decoupled, i.e., $\bar{R}_i = \textrm{diag}\{R_i, R_{j \in \mathcal{N}_i}\}$ and $\frac{1}{2}\bar{u}_i \bar{R}_i \bar{u}_i = \sum\nolimits_{j \in \bar{\mathcal{N}}_i} \frac{1}{2}u_i R_i u_i$. The terminal cost function for the augmented joint state $\bar{Y}_i$ is defined as $\phi_i(\bar{Y}_i) = \sum\nolimits_{j \in \mathcal{N}_i} \omega_j^i \phi_i(x_j) + \omega^i\phi_i(Y_i)$, with $\omega_j^i$ and $ \omega^i > 0$ denoting weights reflecting the importance of each agent. The joint cost-to-go function, subject to the control $\bar{u}_i$ in the first-exit formulation for the augmented subsystem $\bar{\mathcal{N}}_i$, is defined as\begin{equation*}
{\small
    J_i^{\bar{u}_i}(\bar{Y}_i^t, t) = \mathbb{E}_{\bar{Y}_i^t,t}^{\bar{u}_i} [\phi_i(\bar{Y}_i^{t_f}) + \int_t^{t_f} c_i(\bar{Y}_i(\tau), \bar{u}_i(\tau))d\tau],}
\end{equation*}and the joint value function is defined as
\begin{equation}\label{eq:mas-aug-value-function}
\small
    V_i(\bar{Y}_i,t) = \min_{\bar{u}_i}\mathbb{E}_{\bar{Y}_i^t,t}^{\bar{u}_i} [\phi_i(\bar{Y}_i^{t_f}) + \int_t^{t_f} c_i(\bar{Y}_i(\tau), \bar{u}_i(\tau))d\tau].
\end{equation}
Next, we introduce a theorem that summarizes the solution to the optimal control problem for MASs that guarantees safety, which is achieved by solving a linear-form stochastic HJB equation with BaS augmentation.  
\begin{theorem}\label{thm1}
 Consider a MAS consisting of $N$ homogeneous agents with joint dynamics given by~\eqref{eq:cont_dyn_mas}. To incorporate safety, we augment the joint dynamics for subsystem $\bar{\mathcal{N}}_i$ using the central agent BaS, and the augmented joint dynamics are given by~\eqref{eq:mas_aug_dynamics}. The augmented MAS is subject to joint immediate cost~\eqref{eq:mas-aug-immediate-cost} and the joint value function~\eqref{eq:mas-aug-value-function}. Then, the joint optimal control action $\bar{u}_{\mathbf{s}i}^*$ of subsystem $\bar{\mathcal{N}}_i$ ensuring safety is given by
 \begin{equation}\label{eq:mas-aug-joint-ctrl}
     \bar{u}_{\mathbf{s}i}^*(\bar{Y}_i,t) = -\bar{R}_i^{-1} \bar{B}_i^\dagger (\bar{Y}_i)^\top \nabla_{\bar{Y}_i}V_i(\bar{Y}_i,t).
 \end{equation}
 We define the desirability function $Z_i(\bar{Y}_i,t) = \exp[-V_i(\bar{Y}_i,t)/\lambda_i]$, where $\lambda_i \in \mathbb{R}$. Under the safe optimal control action~\eqref{eq:mas-aug-joint-ctrl} and the nonlinearity cancellation condition $\bar{R}_i = (\bar{\sigma}_i \bar{\sigma}_i^\top/\lambda_i)^{-1}$, the joint stochastic HJB equation reduces to a linear form given by
 \begin{align}\label{eq:mas-aug-hjb-eq}
     \partial_t Z_i(\&Y_i,t) &= \big[q_i(\&Y_i)Z_i(\&Y_i,t)/\lambda_i -\&g_i^\dagger (\&Y_i,t)^\top \nabla_{\&Y_i}Z_i(\&Y_i,t)\nonumber \\
     &- \frac{1}{2}\textrm{tr}(\&B_i^\dagger (\&Y_i) \&\sigma_i \&\sigma_i^\top \&B_i^\dagger (\&Y_i)^\top \nabla_{\&Y_i \&Y_i}^2Z_i(\&Y_i,t))\big],
 \end{align}
 with the boundary condition $\&Z_i(\&Y_i,t) = \exp[-\phi_i(\&Y_i)/\lambda_i]$. The above equation can be solved in closed form as \begin{equation}\label{eq:mas-aug-closed-sol}
     Z_i(\&Y_i,t) = \mathbb{E}_{\&Y_i,t}[\exp(-\phi_i(\&p_i^{t_f})/\lambda_i - \int_t^{t_f}q_i(\&p_i)/\lambda_id\tau)].
 \end{equation}
 Here, the diffusion process $\&p_i(t)$ is subject to the uncontrolled dynamics $d\&p_i(\tau) = \&g_i^\dagger(\&p_i,\tau)d\tau + \&B_i^\dagger(\&p_i)\&\sigma_i d\&\omega_i$ with initial condition $\&p_i(t) = \&Y_i(t)$.
\end{theorem}
\begin{remark}
The proof of Theorem~\ref{thm1} is inspired by the proof of Theorem 2 in~\cite{wan2021distributed}, with a few modifications and observations. First, since the central agent states are augmented with the BaSs while other agent states remain unchanged, we need to differentiate the state-dependent terms in the value function, between the central and non-central agents in each subsystem. Second, the optimal control actions now take different forms for central and non-central agents, and therefore the nonlinear term cancellation in the HJB equation needs to be considered separately, with associated optimal control actions. Finally, the safety property of the obtained optimal controls is established by the boundedness of the BaS, which is part of the augmented joint state. A feasible solution to the optimal control problem implies a bounded cost function, and thus bounded BaSs. 
\end{remark}
\begin{proof}
We first substitute the immediate cost function~\eqref{eq:mas-aug-immediate-cost} into~\eqref{eq:mas-aug-value-function}, and denote $s$ as a time step between $t$ and $t_f$; then
\par\nobreak
\vspace{-4mm}
{\small 
  \setlength{\abovedisplayskip}{6pt}
  \setlength{\belowdisplayskip}{\abovedisplayskip}
  \setlength{\abovedisplayshortskip}{0pt}
  \setlength{\belowdisplayshortskip}{3pt}
  \begin{align*} V_i(\&Y_i,t)= 
    \min_{\&u_i}\mathbb{E}_{\&Y_i,t}^{\&u_i}[V_i(\&Y_i,s) &+ \int_t^{s}q_i(\&Y_i) \\
    &+ \frac{1}{2}\&u_i (\&Y_i,\tau)^\top \&R_i \&u_i(\&Y_i,\tau) d\tau], 
\end{align*}}which implies
\par\nobreak
\vspace{-4mm}
{\small 
  \setlength{\abovedisplayskip}{6pt}
  \setlength{\belowdisplayskip}{\abovedisplayskip}
  \setlength{\abovedisplayshortskip}{0pt}
  \setlength{\belowdisplayshortskip}{3pt}
  \begin{align*} 
   0 = \min_{\&u_i}&\mathbb{E}_{\&Y_i,t}^{\&u_i}\big[(V_i(\&Y_i,s)-V_i(\&Y_i,t))/(s-t) \\
    &+ \frac{1}{s-t}\int_t^{s}q_i(\&Y_i) 
    + \frac{1}{2}\&u_i (\&Y_i,\tau)^\top \&R_i \&u_i(\&Y_i,\tau) d\tau\big].
  \end{align*}}By letting $s \rightarrow t$, the optimality equation takes the form
\begin{equation}\label{eq:mas-aug-optimality-proof}
\small
    0 = \min_{\&u_i}\mathbb{E}_{\&Y_i,t}^{\&u_i}\big[\frac{dV_i(\&Y_i,t)}{dt} 
    + q_i(\&Y_i) 
    + \frac{1}{2}\&u_i (\&Y_i,t)^\top \&R_i \&u_i(\&Y_i,t) \big].
\end{equation}
By expanding the $dV_i(\&Y_i,t)/dt$ term using It$\hat{\textrm{o}}$'s formula,  
and taking the expectation over all trajectories that initialized at $(\&Y_i^t,t)$ and subject to control $\&u_i$, we have  
\par\nobreak
\vspace{-4mm}
{\small 
  \setlength{\abovedisplayskip}{6pt}
  \setlength{\belowdisplayskip}{\abovedisplayskip}
  \setlength{\abovedisplayshortskip}{0pt}
  \setlength{\belowdisplayshortskip}{3pt}
  \begin{align}\label{eq:mas-aug-expectation-expanded-proof}
   &\mathbb{E}_{\&Y_i,t}^{\&u_i} [dV_i(\&Y_i,t)/dt] = \partial V_i(\&Y_i,t)/\partial t \nonumber\\
    &+ \sum\nolimits_{j \in \mathcal{N}_i}[g_j(x_j,t)+B_j(x_j)u_j(\&Y_i,t)]^\top \nabla_{x_j}V_i(\&Y_i,t)\nonumber \\
    &+ [g_i^\dagger(Y_i,t)+B_i^\dagger(Y_i)u_i(\&Y_i,t)]^\top \nabla_{Y_i}V_i(\&Y_i,t)\nonumber\\
    &+ \frac{1}{2}\textrm{tr}(B_i^\dagger(Y_i) \sigma_i \sigma_i^\top B_i^\dagger(Y_i)^\top \nabla_{Y_iY_i}^2V_i(\&Y_i,t))\nonumber\\
    &+ \frac{1}{2}\sum\nolimits_{j \in \mathcal{N}_i}\textrm{tr}(B_j(x_j) \sigma_j \sigma_j^\top B_j(x_j)^\top \nabla_{x_jx_j}^2V_i(\&Y_i,t)).
\end{align}}
Here, we use several identities: $\mathbb{E}_{\&Y_i,t}^{\&u_i}[dY_{i(m)}dx_{j(n)}] = (\sigma_i \sigma_i^\top)_{mm}\delta_{ij} \delta_{mn} dt$, $\mathbb{E}_{\&Y_i,t}^{\&u_i}[dY_{i(m)}dY_{i(n)}] = (\sigma_i \sigma_i^\top)_{mm} \delta_{mn} dt$, $\mathbb{E}_{\&Y_i,t}^{\&u_i}[dx_{j(m)}dx_{j(n)}] = (\sigma_j \sigma_j^\top)_{mm} \delta_{mn} dt$,
where $(\cdot)_{mm}$ denotes the $(m,m)$ entry of one matrix, and $\delta_{ij} = 1 (0)$ if $i=j (i \ne j)$. Since $j \in \mathcal{N}_i$ and $j \ne i$, we have $ \mathbb{E}_{\&Y_i,t}^{\&u_i}[dY_{i(m)}dx_{j(n)}] = 0$. Substituting~\eqref{eq:mas-aug-expectation-expanded-proof} into~\eqref{eq:mas-aug-optimality-proof} with above identities, the joint stochastic HJB equation for the augmented system is obtained as
\par\nobreak
\vspace{-4mm}
{\small 
  \setlength{\abovedisplayskip}{6pt}
  \setlength{\belowdisplayskip}{\abovedisplayskip}
  \setlength{\abovedisplayshortskip}{0pt}
  \setlength{\belowdisplayshortskip}{3pt}
  \begin{align}\label{eq:mas-aug-stochastic-hjb-simplified-proof}
     &-\partial_t V_i(\&Y_i,t) = \nonumber\\
    &\min_{\&u_i} \mathbb{E}_{\&Y_i,t}^{\&u_i}\big[[g_i^\dagger(Y_i,t)+B_i^\dagger(Y_i)u_i(\&Y_i,t)]^\top \nabla_{Y_i}V_i(\&Y_i,t)\nonumber\\
    &+ \sum\nolimits_{j \in \mathcal{N}_i} [g_j(x_j,t) + B_j(x_j)u_j(\&Y_i,t)]^\top \nabla_{x_j}V_i(\&Y_i,t)\nonumber\\
    &+ q_i(\&Y_i) + \frac{1}{2}\&u_i(\&Y_i,t)^\top \&R_i \&u_i(\&Y_i,t)\nonumber\\
    &+ \frac{1}{2}\textrm{tr}(B_i^\dagger(Y_i)\sigma_i \sigma_i^\top B_i^\dagger (Y_i)^\top \nabla_{Y_iY_i}^2 V_i(\&Y_i,t))\nonumber\\
    &+ \frac{1}{2}\sum\nolimits_{j\in \mathcal{N}_i}\textrm{tr}(B_j(x_j)\sigma_j \sigma_j^\top B_j (x_j)^\top \nabla_{x_jx_j}^2 V_i(\&Y_i,t))\big],
\end{align}}where the boundary condition is given by $V_i(\&Y_i,t_f) = \phi_i(\&Y_i)$. Since the RHS of~\eqref{eq:mas-aug-stochastic-hjb-simplified-proof} is quadratic in control, the safe optimal control action can be obtained by setting the derivatives of the operand with respect to $u_j(\&Y_i,t)$ and $u_i(\&Y_i,t)$ equal to zero, respectively, and we have
\par\nobreak
\vspace{-1em}
{\small 
  \setlength{\abovedisplayskip}{6pt}
  \setlength{\belowdisplayskip}{\abovedisplayskip}
  \setlength{\abovedisplayshortskip}{0pt}
  \setlength{\belowdisplayshortskip}{3pt}
  \begin{align}\label{eq:mas-aug-opt-ctrl-central-and-noncentral-proof}
   u_{\mathbf{s}i}^*(\&Y_i,t) &= -R_i^{-1}B_i^\dagger(Y_i)^\top \nabla_{Y_i}V_i(\&Y_i,t), \nonumber\\ u_{\mathbf{s}j}^*(\&Y_i,t) &= -R_j^{-1}B_j(x_j)^\top \nabla_{x_j}V_i(\&Y_i,t).
\end{align}}With the exponential transformation of value function $Z_i(\&Y_i,t) = \exp[-V_i(\&Y_i,t)/\lambda_i]$, we can represent the terms involving $V_i(\&Y_i,t)$ in~\eqref{eq:mas-aug-stochastic-hjb-simplified-proof} and~\eqref{eq:mas-aug-opt-ctrl-central-and-noncentral-proof} with respect to $Z_i(\&Y_i,t)$.
For each non-central agent $j \in \mathcal{N}_i$, by adopting the optimal control action $u_{\mathbf{s}j}^*$, and according to the property of trace operator,  we can simplify the terms in~\eqref{eq:mas-aug-stochastic-hjb-simplified-proof} as follows:
\par\nobreak
\vspace{-4mm}
{\small 
  \setlength{\abovedisplayskip}{6pt}
  \setlength{\belowdisplayskip}{\abovedisplayskip}
  \setlength{\abovedisplayshortskip}{0pt}
  \setlength{\belowdisplayshortskip}{3pt}
  \begin{align}\label{eq:mas-aug-stochastic-hjb-term1}
    &[B_j(x_j)u_j(\&Y_i,t)]^\top \nabla_{x_j}V_i(\&Y_i,t) + \frac{1}{2}u_j(\&Y_i,t)^\top R_ju_j(\&Y_i,t)\nonumber\\
    &= -\frac{\lambda_i^2}{2}\frac{\nabla_{x_j}Z_i(\&Y_i,t)^\top B_j(x_j)R_j^{-1}B_j(x_j)^\top \nabla_{x_j}Z_i(\&Y_i,t)}{Z_i^2(\&Y_i,t)},
\end{align}}
\par\nobreak
\vspace{-4mm}
{\small 
  \setlength{\abovedisplayskip}{6pt}
  \setlength{\belowdisplayskip}{\abovedisplayskip}
  \setlength{\abovedisplayshortskip}{0pt}
  \setlength{\belowdisplayshortskip}{3pt}
  \begin{align}\label{eq:mas-aug-stochastic-hjb-term2}
  &\frac{1}{2}\textrm{tr}(B_j(x_j)\sigma_j\sigma_j^\top B_j(x_j)^\top \nabla_{x_jx_j}^2V_i(\&Y_i,t))\nonumber\\
    &= -\lambda_i \textrm{tr}(B_j(x_j)\sigma_j\sigma_j^\top B_j(x_j)^\top \nabla_{x_jx_j}^2Z_i(\&Y_i,t))/2Z_i(\&Y_i,t)\nonumber \\
    &+ \frac{\lambda_i \textrm{tr}(\nabla_{x_j}Z_i(\&Y_i,t)^\top B_j(x_j)\sigma_j\sigma_j^\top B_j(x_j)^\top \nabla_{x_j}Z_i(\&Y_i,t))}{2Z_i^2(\&Y_i,t)}.
\end{align}}
The quadratic terms on the desirability gradients in~\eqref{eq:mas-aug-stochastic-hjb-term1} and~\eqref{eq:mas-aug-stochastic-hjb-term2} are cancelled by selecting $\sigma_j\sigma_j^\top = \lambda_i R_j^{-1}$. Similarly, the quadratic terms with respect to the desirability gradients involving central agent $i$ in~\eqref{eq:mas-aug-stochastic-hjb-simplified-proof} are cancelled by selecting $\sigma_i\sigma_i^\top = \lambda_i R_i^{-1}$, i.e., the nonlinearity cancellation condition can be further rewritten as $\&\sigma_i\&\sigma_i^\top = \lambda_i\&R_i^{-1}$, and the linear-form stochastic HJB equation is thus obtained
\par\nobreak
\vspace{-4mm}
{\small 
  \setlength{\abovedisplayskip}{6pt}
  \setlength{\belowdisplayskip}{\abovedisplayskip}
  \setlength{\abovedisplayshortskip}{0pt}
  \setlength{\belowdisplayshortskip}{3pt}
  \begin{align}\label{eq:mas-aug-linear-stochastic-hjb-ultimate-proof}
     \partial_t Z_i(\&Y_i,t) &= \big[\frac{q_i(\&Y_i)Z_i(\&Y_i,t)}{\lambda_i} - \&g_i^\dagger(\&Y_i,t)^\top \nabla_{\&Y_i}Z_i(\&Y_i,t) \nonumber \\
    &-\frac{1}{2}\textrm{tr}(\&B_i^\dagger(\&Y_i)\&\sigma_i\&\sigma_i^\top \&B_i^\dagger(\&Y_i)^\top \nabla_{\&Y_i\&Y_i}^2Z_i(\&Y_i,t))\big],
\end{align}}where the boundary condition is given by $Z_i(\&Y_i,t_f) = \exp[-\phi_i(\&Y_i)/\lambda_i]$. By invoking the Feynman-Kac formula, a solution to~\eqref{eq:mas-aug-linear-stochastic-hjb-ultimate-proof} can be formulated as\begin{equation*}
\small
    Z_i(\&Y_i,t) = \mathbb{E}_{\&Y_i,t}\big[\exp(-\phi_i(\&p_i^{t_f})/\lambda_i-\int_t^{t_f}q_i(\&p_i)/\lambda_id\tau)\big],
\end{equation*}
where diffusion process $\&p(t)$ satisfies $d\&p_i(t) = \&g_i^\dagger(\&p_i,\tau)d\tau + \&B_i^\dagger(\&p_i)\&\sigma_id\&\omega_i$ with the initial condition $\&p_i(t) = \&Y_i(t)$.
\end{proof}
\subsection{Path integral approximation formulation}
Although Theorem~1 of Section~\ref{sec:aug_opt_ctrl_mas} provides a closed-form solution of the optimal control action, computing the expectation term over all trajectories initiated at $(\bar{Y}_i^t, t)$ in the solution of the desirability function~\eqref{eq:mas-aug-closed-sol} is usually intractable. To address this, we reformulate the optimal control solution for the augmented system using path integrals and propose a revised proposition based on Proposition 3 in~\textcolor{black}{~\cite{wan2021distributed}}.
\begin{proposition}\label{prop1}
Based on the partition of augmented joint state $\bar{Y}_i$ into directly actuated states and non-directly actuated states as introduced in~\eqref{eq:mas_aug_dynamics_partitioned}, we can further partition the time interval $[t,t_f]$ into $K$ equal-length ($\varepsilon = (t_f-t)/K$) intervals, i.e., $t = t_0 < t_1 < t_2 < \dots < t_K = t_f$, and use the trajectory variables $\&Y_i^{(k)} = [\&Y_{i(n)}^{(k)^\top},\&Y_{i(d)}^{(k)^\top}]^\top$ to denote the joint uncontrolled ($\&u_i(\&Y_i,t) = 0$) trajectories of the augmented system on time $[t_{k-1},t_k)$ with initial condition $\&Y_i(t) = \&Y_i^{(0)}$. We use $\&l_i = (\&Y_i^{(1)}, \&Y_i^{(2)}, \dots, \&Y_i^{(K)})$ to denote the path variable, and the generalized path value is formulated as
\par\nobreak
\vspace{-4mm}
{\small
  \setlength{\abovedisplayskip}{6pt}
  \setlength{\belowdisplayskip}{\abovedisplayskip}
  \setlength{\abovedisplayshortskip}{0pt}
  \setlength{\belowdisplayshortskip}{3pt}
  \begin{align}\label{eq:mas_aug_generalized-pv}
    &\tilde{S}_i^{\varepsilon,\lambda_i}(\&Y_i^{(0)},\&l_i,t_0) = \frac{\phi_i(\&Y_i^{(k)})}{\lambda_i} + \frac{ \varepsilon}{\lambda_i}\sum\nolimits_{k=0}^{K-1}q_i(\&Y_i^{(k)}) \nonumber\\
   &+ \frac{1}{2}\sum\nolimits_{k=0}^{K-1} \log\det(H_i^{(k)}) + \frac{ \varepsilon}{2\lambda_i}\sum\nolimits_{k=0}^{K-1}\norm[\big]{\alpha_i^{(k)}}^2_{(H_i^{(k)})^{-1}},
\end{align}}with $H_i^{(k)} = \&B_{i(d)}^\dagger (\&Y_i^{(k)})\&\sigma_i \&\sigma_i^\top \&B_{i(d)}^\dagger (\&Y_i^{(k)})^\top$ and $\alpha_i^{(k)} = (\&Y_{i(d)}^{(k+1)} - \&Y_{i(d)}^{(k)})/\varepsilon - \&g_{i(d)}^\dagger(\&Y_i^{(k)},t_k) $. Then, the joint optimal control action ensuring safety in subsystem $\bar{\mathcal{N}}_i$ in~\eqref{eq:mas-aug-joint-ctrl} can be formulated as path integral
\par\nobreak
\vspace{-4mm}
{\small 
  \setlength{\abovedisplayskip}{6pt}
  \setlength{\belowdisplayskip}{\abovedisplayskip}
  \setlength{\abovedisplayshortskip}{0pt}
  \setlength{\belowdisplayshortskip}{3pt}
  \begin{align*}
    &\&u_{\mathbf{s}i}^*(\&Y_i,t) = \nonumber \\ &\lambda_i \&R_i^{-1}\&B_{i(d)}^\dagger (\&Y_i)^\top \cdot \lim_{\varepsilon \downarrow 0}\int \tilde{p}_i^*(\&l_i|\&Y_i^{(0)},t_0)\cdot \tilde{u}_i(\&Y_i^{(0)},\&l_i,t_0) d\&l_i,
\end{align*}}where\begin{equation*}
{\small
    \tilde{p}_i^*(\&l_i|\&Y_i^{(0)},t_0) =
 \frac{\exp(-\tilde{S}_i^{\varepsilon,\lambda_i}(\&Y_i^{(0)},\&l_i,t_0))}{\int \exp(-\tilde{S}_i^{\varepsilon,\lambda_i}(\&Y_i^{(0)},\&l_i,t_0)) d\&l_i}}
\end{equation*}
is the optimal path distribution and 
\par\nobreak
\vspace{-4mm}
{\small
  \setlength{\abovedisplayskip}{6pt}
  \setlength{\belowdisplayskip}{\abovedisplayskip}
  \setlength{\abovedisplayshortskip}{0pt}
  \setlength{\belowdisplayshortskip}{3pt}
  \begin{align*}
    \tilde{u}_i(\&Y_i^{(0)},\&l_i,t_0) &= -\frac{\varepsilon}{\lambda_i}\nabla_{\&Y_{i(d)}^{(0)}}q_i(\&Y_i^{(0)})\\*
    &+ (H_i^{(0)})^{-1}\big((\&Y_{i(d)}^{(1)}-\&Y_{i(d)}^{(0)})/\varepsilon - \&g_{i(d)}^\dagger (\&Y_i^{(0)},t_0)\big)
\end{align*}}is the initial control variable.
\end{proposition}
\begin{remark}
Once feasible solutions to the cost-minimization problem (i.e.,~\eqref{eq:mas-aug-value-function}) for the BaS-augmented subsystems are determined, which is expressed in a closed form in Theorem~\ref{thm1} and approximated in Proposition~\ref{prop1}, the cost function must be bounded. Since the BaSs are also part of the cost function, the resulting BaSs are bounded. (Rigorously, the BaS boundedness is established in a mean-square sense, and the achieved safety is also in a mean-square sense). According to Lemma~\ref{lemma1} and Remark~\ref{remark_bas}, the achieved optimal control action to the augmented subsystem ensures safety of the central agent within the original subsystem. Taking safe optimal control actions for agent $i$ from individual subsystem $\bar{\mathcal{N}}_i$ then collectively ensures safety of the entire network.
\end{remark}
\section{SIMULATION RESULTS}\label{simulation}
In this section, we conduct numerical simulations on cooperative MASs consisting of UAVs in environments with obstacles. The objective is to reach targets, avoid obstacles, and cooperate with other agents. The continuous-time dynamics of each UAV are given by:\begin{equation}\label{eq:simulation-uav-dynamics}
    \setlength\arraycolsep{1pt}
	{\scalefont{0.8} \left(\begin{matrix}
		dx_i\\
		dy_i\\
		dv_i\\
		d\varphi_i
		\end{matrix}\right) = 
		\left(\begin{matrix}
		v_i \cos \varphi_i\\
		v_i \sin \varphi_i\\
		0\\
		0
		\end{matrix}\right) dt +  \begin{pmatrix}
		0 & 0\\
		0 & 0\\
		1 & 0\\
		0 & 1
		\end{pmatrix} \left[ \left( \begin{matrix}
		u_i\\
		w_i
		\end{matrix}  \right) dt + \begin{pmatrix}
		\sigma_i & 0\\
		0 & \nu_i
		\end{pmatrix}d\omega_i
		\right], }
\end{equation}where $(x_i,y_i),v_i,\varphi_i$ represent the position coordinate, forward velocity, and heading angle of UAV $i$. We use $\mathbf{x}_i := (x_i,y_i,v_i,\varphi_i)^\top$ to denote the state vector. The forward acceleration $u_i$ and angular velocity $w_i$ are the control inputs, and $\omega_i$ is the standard Brownian motion disturbance. We set the noise level to $\sigma_i = 0.1$ and $\nu_i = 0.05$, and specify the exit time as $t_f=20$ seconds. 

\begin{figure}[!h]
\centerline{\includegraphics[width=0.6\columnwidth]{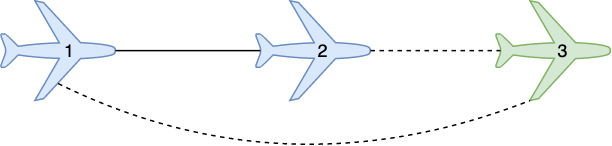}}
\caption{A networked UAV team with UAVs 1 and 2 flying cooperatively and UAV 3 flying independently.}
\label{fig_uav_network}
\end{figure}

In the following examples, we utilize the safe optimal control action~\eqref{eq:mas-aug-joint-ctrl} introduced in Theorem 1, which we approximate using the path-integral formulation discussed in Proposition~\ref{prop1}. Specifically, we sample local safe optimal control actions for each central agent from the computed joint control actions in each subsystem. To tackle obstacle-avoidance tasks, we first model all the safety constraints with independent BaSs and then augment the joint dynamics for subsystems with the central agent BaSs' dynamics. 
\subsection{Passing through simple obstacles}

In this example, we consider a cooperative UAV team as depicted in Figure~\ref{fig_uav_network}, where UAVs 1 and 2 fly together with their distance minimized and UAV 3 operates independently. The dotted lines in Figure~\ref{fig_uav_network} indicate that the coordination between UAVs is not factored into the running cost function design, resulting in a loosely-connected team. Conversely, UAVs connected by a solid line can cooperate with one another. Each UAV is governed by the continuous-time dynamics~\eqref{eq:simulation-uav-dynamics}. The team of three UAVs is tasked to navigate through obstacles, flying from $(5,5),(5,45)$, and $(5,25)$ to $(45,25)$, while avoiding all obstacles. The three circular obstacles in the environment are described by the following three functions: $h_1(\mathbf{x}) = (x-17)^2 + (y-40)^2 -8^2$, $h_2(\mathbf{x}) = (x-22)^2 + (y-16)^2 - 7^2$, and $h_3(\mathbf{x}) = (x-35)^2 + (y-30)^2 - 5^2$. The safe region is represented as the intersection of the safe sets, i.e., $\bar{\mathcal{C}} = \cap_{i=1}^3 \mathcal{C}_i$ with $\mathcal{C}_i := \{\mathbf{x}\in \mathbb{R}^2: h_i(\mathbf{x}) > 0\}$. We adopt independent BaSs to represent the constraints for each subsystem's central agent, where $z_{ij}$ is the BaS corresponding to constraint $j$ for agent $i$, generated by~\eqref{eq:sas_single_bas_dynamics_simplified}. As discussed in Lemma~\ref{lemma1}, we initialize $z_{ij}(0)$ to $ \beta_j(\mathbf{x}_i(0))-\beta_j(0)$, where $\beta_j(\cdot)$ is the inverse BF. The $\gamma$-parameter in~\eqref{eq:sas_single_bas_dynamics_simplified} is set to be 0.5, which determines the BaS's rate of returning to $\beta(\mathbf{x}_i)-\beta_0$, as discussed in~\cite{almubarak2021safety}.

We assume that the communication network of the UAV team is fully-connected, i.e., all the UAVs can sense the states of other UAVs. Based on the factorization topology introduced in Section~\ref{sec:distributed-network-framework}, the joint states of the three factorial subsystems are: $\&{\mathbf{x}}_1 = [\mathbf{x}_1,\mathbf{x}_2,\mathbf{x}_3]^\top; \&{\mathbf{x}}_2 = [\mathbf{x}_1,\mathbf{x}_2,\mathbf{x}_3]^\top; \&{\mathbf{x}}_3 = [\mathbf{x}_1,\mathbf{x}_2,\mathbf{x}_3]^\top$, where $\mathbf{x}_i = [x_i,y_i,v_i,\varphi_i]^\top$ is the state variable vector. The running cost functions for the three subsystems are designed as $ q(\&{\mathbf{x}}_1) = 3.5(\|(x_1,y_1)-(x_1^{t_f},y_1^{t_f})\|_2 - d_1^{\max}) + 1.4(\|(x_1,y_1)-(x_2,y_2)\|_2 - d_{12}^{\max}) +  0.5\|(z_{11},z_{12},z_{13})-(z_{11}^{t_f},z_{12}^{t_f},z_{13}^{t_f})\|_2^2 + 50\prod\nolimits_{j=1}^3 \mathbb{I}(z_{1j} > 0.01), \hspace{2pt} q(\&{\mathbf{x}}_2) = 3.5(\|(x_2,y_2)-(x_2^{t_f},y_2^{t_f})\|_2 - d_2^{\max})+ 1.4(\|(x_2,y_2)-(x_1,y_1)\|_2 - d_{21}^{\max})+  1.5\|(z_{21},z_{22},z_{23})-(z_{21}^{t_f},z_{22}^{t_f},z_{23}^{t_f})\|_2^2 + 50\prod\nolimits_{j=1}^3 \mathbb{I}(z_{2j} > 0.01), q(\&{\mathbf{x}}_3) = 6(\|(x_3,y_3)-(x_3^{t_f},y_3^{t_f})\|_2 - d_3^{\max})+  0.5\|(z_{31},z_{32},z_{33})-(z_{31}^{t_f},z_{32}^{t_f},z_{33}^{t_f})\|_2^2 + 50\prod\nolimits_{j=1}^3 \mathbb{I}(z_{3j} > 0.01)$,
where $\|(x_i,y_i)-(x_i^{t_f},y_i^{t_f})\|$ computes the distance to the target for UAV $i$, $\|(x_i,y_i)-(x_j,y_j)\|$ computes the distance between UAV $i$ and $j$. Here, the hyper-parameter $d_i^{\max}$ represents the maximum difference between the initial position of UAV $i$ and its target. Similarly, $d_{ij}^{\max}$ represents the initial distance between UAVs $i$ and $j$. Besides the BaS dependency in running cost functions, large BaSs are also penalized when the indicator function $\mathbb{I}(\cdot)$ is activated. The simulation is performed using a step size of $\Delta t = 0.05s$.

\begin{figure}[!h]
\centerline{\includegraphics[width=0.8\columnwidth]{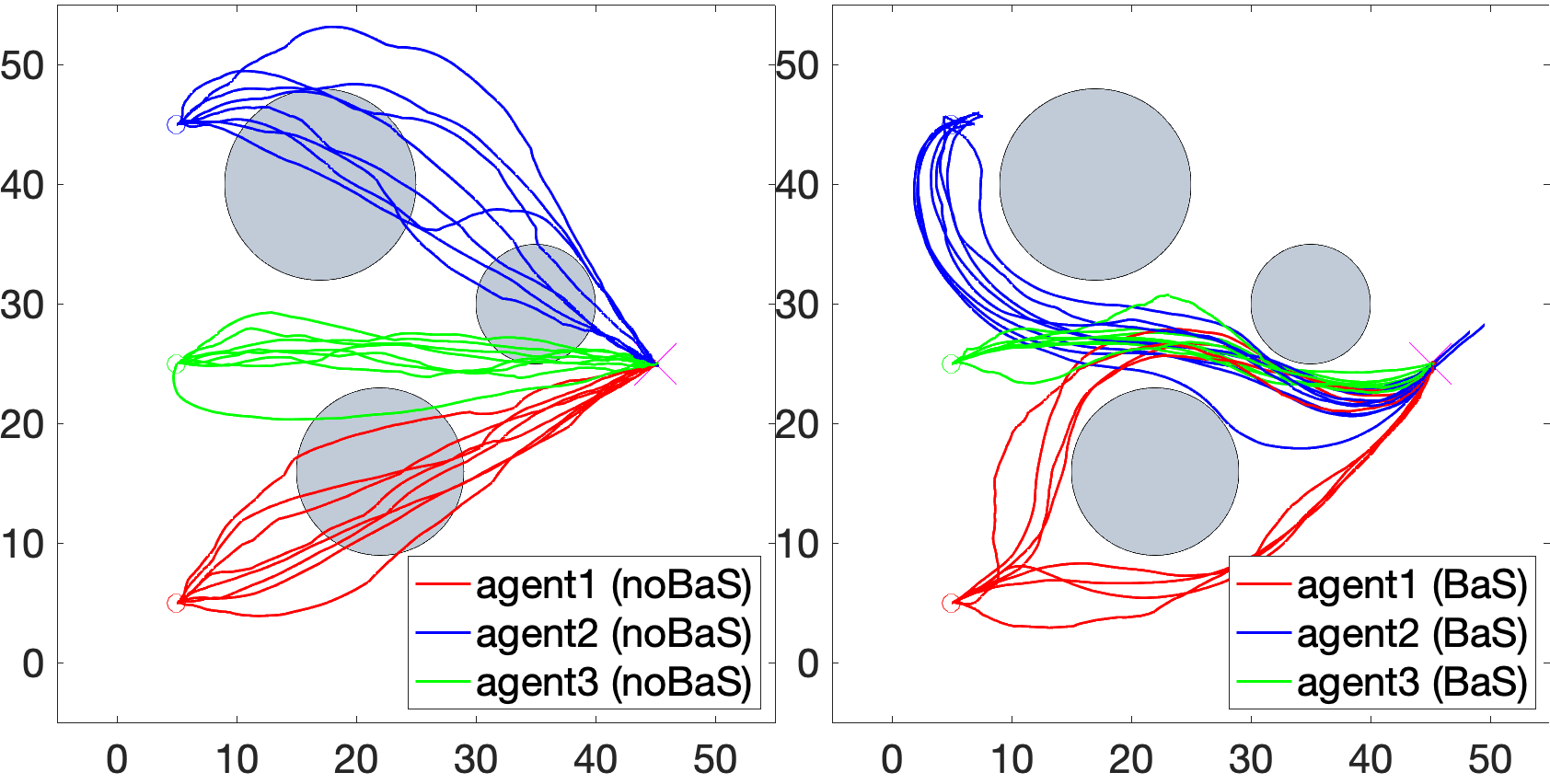}}
\caption{Comparison between executed trajectories with original optimal control (left) and BaS-augmented safe optimal control (right) on a UAV team in environment with obstacles.}
\label{fig_comp_multi_agent_new}
\end{figure}

We compare the safe optimal control via the BaS augmentation method with the original optimal control method, which only includes constraint-violation penalty in the cost function. We run 8 simulations with each method under identical initial conditions on the UAV team, and the result is illustrated in Figure~\ref{fig_comp_multi_agent_new}. We denote the trajectories for UAVs 1,2, and 3 with red, blue, and green lines, respectively. The starting points for all UAVs are denoted by small circles, and the target is represented by a cross mark. Both methods successfully drive the UAV team to the target. However, most of the executed trajectories of the original optimal control still collide with the surrounding obstacles, and safety is not guaranteed, while the proposed method always ensures safety by avoiding all the obstacles. We design UAVs 1 and 2 to fly together as one of our control objectives. In Figure~\ref{fig_comp_multi_agent_new}, we observe UAVs 1 (red line) and 2 (blue line) make attempt to minimize the distance between them with the proposed method, which is not achieved with the original optimal control method. The better coordination performance can be explained as the obstacle-collision penalty (usually large for safety concerns) in the cost function of original optimal control entangles the agent coordination objective. However, with the BaS-augmented safe optimal control, which essentially encodes the safety constraints into BaSs without introducing explicit obstacle-collision penalty terms, other control objectives are not sacrificed.  

\subsection{Passing through cluttered-obstacles}
 We further consider the same cooperative UAV team passing through a more obstacle-cluttered environment. The team of three UAVs is tasked to navigate through an obstacle-cluttered environment, flying from $(2.5,2.5), (2.5, 70)$, $(2.5,40)$ to $(45,45)$, while avoiding all obstacles. The five circular obstacles in the environment are described by the following five functions: $h_1(\mathbf{x}) = (x-15)^2 + (y-3)^2 -8^2$, $h_2(\mathbf{x}) = (x-16)^2 + (y-27)^2 -8^2$,  $h_3(\mathbf{x}) = (x-27)^2 + (y-15)^2 -6^2$, $h_4(\mathbf{x}) = (x-32.5)^2 + (y-30.5)^2 -12^2$, and $h_5(\mathbf{x}) = (x-20)^2 + (y-60)^2 -4^2$. The safe region is represented as the intersection of the safe sets, i.e.,  $\bar{\mathcal{C}} = \cap_{i=1}^5 \mathcal{C}_i$ with $\mathcal{C}_i := \{\mathbf{x}\in \mathbb{R}^2: h_i(\mathbf{x}) > 0\}$. We adopt the same $\gamma$ and BaS-construction methodology as in the first experiment. The running cost functions for the three subsystems are designed as follows: $ q(\&{\mathbf{x}}_1) = 2(\|(x_1,y_1)-(x_1^{t_f},y_1^{t_f})\|_2 - d_1^{\max}) + 0.5(\|(x_1,y_1)-(x_2,y_2)\|_2 - d_{12}^{\max}) +  0.8\|(z_{11},z_{12},z_{13},z_{14},z_{15})-(z_{11}^{t_f},z_{12}^{t_f},z_{13}^{t_f},z_{14}^{t_f},z_{15}^{t_f})\|_2^2 + 24\prod\nolimits_{j=1}^5 \mathbb{I}(z_{1j} > 0.01), \hspace{2pt} q(\&{\mathbf{x}}_2) = 2(\|(x_2,y_2)-(x_2^{t_f},y_2^{t_f})\|_2 - d_2^{\max}) + 0.5(\|(x_2,y_2)-(x_1,y_1)\|_2 - d_{21}^{\max}) +  0.8\|(z_{21},z_{22},z_{23},z_{24},z_{25})-(z_{21}^{t_f},z_{22}^{t_f},z_{23}^{t_f},z_{24}^{t_f},z_{25}^{t_f})\|_2^2 + 24\prod\nolimits_{j=1}^5 \mathbb{I}(z_{2j} > 0.01), q(\&{\mathbf{x}}_3) = 3(\|(x_3,y_3)-(x_3^{t_f},y_3^{t_f})\|_2 - d_3^{\max})+  0.8\|(z_{31},z_{32},z_{33},z_{34},z_{35})-(z_{31}^{t_f},z_{32}^{t_f},z_{33}^{t_f},z_{34}^{t_f},z_{35}^{t_f})\|_2^2 + 48\prod\nolimits_{j=1}^5 \mathbb{I}(z_{3j} > 0.01)$. Here, all the hyper-parameters (e.g., $d_1^{\max}$) follow the same definitions as in the first experiment. The simulation is performed using a step size of $\Delta t = 0.2s$.
 
\begin{figure}
\centerline{\includegraphics[width=0.9\columnwidth]{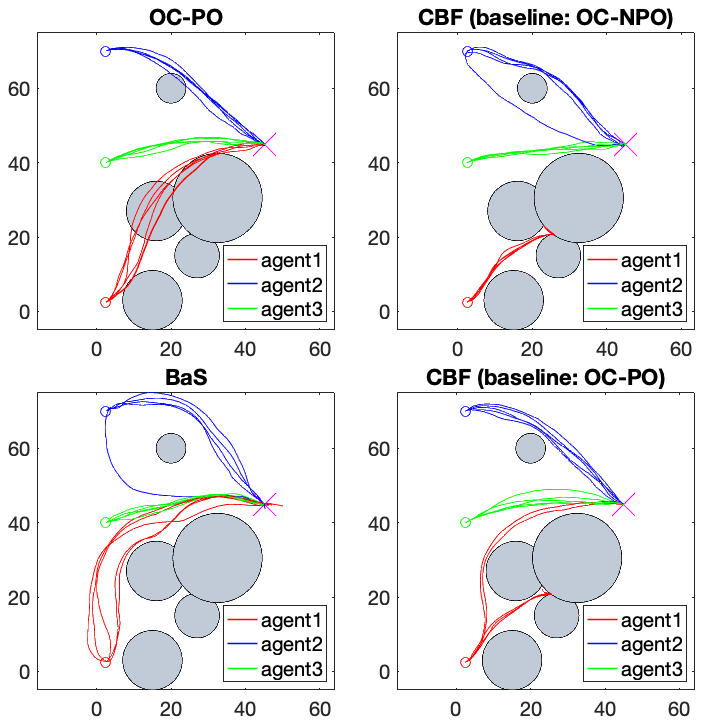}}
\caption{Comparison of executed trajectories under optimal control penalizing obstacle-collision (top-left, abbreviation: `\textbf{OC-PO}'), CBF plus baseline optimal control without penalizing obstacle-collision (top-right, abbreviation: `\textbf{CBF (baseline: OC-NPO)}'), CBF plus baseline optimal control penalizing obstacle-collision (bottom-right, abbreviation: `\textbf{CBF (baseline: OC-PO)}'), and BaS-augmented optimal control (bottom-left) for a UAV team in an obstacle-cluttered environment.}
\label{fig_comp_multi_agent_comparison_new_final}
\end{figure}

We employ four different approaches to compute optimal control solutions for the goal-reaching, agent-coordinating, and obstacle-avoidance task. In the conventional optimal control framework, any violation of the safety constraints is penalized in the cost function. The CBF-based safe optimal controller, discussed in~\cite{song2022generalization,clark2019control}, first computes a baseline optimal control and then uses CBF as a safety filter to modify the baseline control input if necessary to ensure that the safety constraints are satisfied. Here, we categorize the CBF-based safe optimal controller based on whether the obstacle-collision penalty is included in the baseline control cost function design. We compare these three methods with the BaS-based safe optimal control method proposed in this paper, which solves optimal control action on the joint dynamics with augmentation of the central agent BaSs. We conduct five simulations with each method under identical initial conditions, and the results are compared in Figure~\ref{fig_comp_multi_agent_comparison_new_final}.

From Figure~\ref{fig_comp_multi_agent_comparison_new_final}, it is observed that the original optimal control solution obtained by solely penalizing constraint violation is insufficient to avoid obstacles in a cluttered environment. Additionally, the CBF-based method, which involves filtering the baseline optimal control by solving a quadratic programming (QP) problem, got trapped in obstacles and failed to reach the target in some scenarios. This behavior is essentially due to the fact that CBF-based methods are reactive to given safety constraints and require good tuning to ensure feasibility and task completion. In contrast, the proposed safe optimal control method using BaS augmentation ensures safety while successfully reaching the target and achieving desired coordination in all trails. 

\section{CONCLUSION AND FUTURE WORK}\label{conclusion}
In this paper, we introduce a novel safety-embedded stochastic optimal control framework for networked multi-agent systems (MASs) using barrier states (BaSs) augmentation.  Our approach involves augmenting the joint dynamics of each factorial subsystem by introducing BaSs that embed safety constraints for the central agent and solving the optimal control action on the augmented subsystem. The proposed method simultaneously achieves safety and other control objectives, where safety is guaranteed by the boundedness of BaSs and achieved by the feasible solutions to the reformulated optimal control problem. The safe optimal control law is obtained by solving a linear stochastic Hamilton-Jacobi-Bellman (HJB) equation, and the solution is approximated using the path-integral formulation. We validate our approach through numerical simulations on a networked team of UAV. Future work includes representing the optimal control solution in policy improvement with path integrals (PI\textsuperscript{2}) framework and exploring other approximation formulations, such as Relative Entropy Policy Search (REPS). Moreover, we aim to extend our safety-embedded stochastic optimal control framework to MASs with incomplete state.

\section{ACKNOWLEDGMENT}
The authors would like to appreciate the constructive discussions with Hassan Almubarak.

\addtolength{\textheight}{-12cm}   

\bibliography{reference}
\bibliographystyle{IEEEtran}
\end{document}